\documentclass[aps,twocolumn,PRL,reprint,showpacs,superscriptaddress]{revtex4}
\usepackage{amsmath}
\usepackage[percent]{overpic}
\usepackage{times}
\usepackage{subfigure}
\usepackage{latexsym}
\usepackage{graphicx}
\usepackage{bm}
\usepackage{amssymb}
\usepackage{hyperref}
\usepackage{multirow}

\begin{document}

%\preprint{APS/123-QED}

\title{Surface excitations, shape deformation and the long-time behavior in a stirred Bose-Einstein condensate}

\author{Qing-Li Zhu}
\affiliation{National Laboratory of Solid State Microstructures and Department of Physics, Nanjing University, Nanjing 210093, China}
\author{Jin An}
\email{anjin@nju.edu.cn}
\affiliation{National Laboratory of Solid State Microstructures and Department of Physics, Nanjing University, Nanjing 210093, China}
\affiliation{Collaborative Innovation Center of Advanced Microstructures, Nanjing University, Nanjing 210093, China}
\date{\today}

\begin{abstract}
The surface excitations, shape deformation and the formation of persistent current for a Gaussian obstacle potential rotating in an highly oblate
Bose-Einstein condensate(BEC)are investigated. Vortex dipole can be produced and trapped in the center of the stirrer even for slow motion of
the stirring beam. When the barrier angular velocity is above some critical value, the condensate shape can be deformed remarkably according to the
rotation frequency due to the existence of plenty of surface wave excitations. After a long enough time, a few vortices are found to be left either
trapped in the condensate or pinned by the obstacle, a vortex dipole or several vortices can be trapped at the beam center, which enables the possibility
of vortex manipulation.
\end{abstract}
\maketitle

\emph{Introduction.}---Quantized vortex is a topological singularity in a superfluid or superconductor, where the phase of the order parameter varies by integral multiples of $2\pi$ whenfollowing a closed path around the defect. Due to its topological nature and the conservation of circulation, a vortex can be eliminated only by annihilation with an
antivortex or moving to the boundary of system.

Trapped Bose-Einstein Condensate(BEC) of ultracold atomic gases provides a very convenient platform to investigate the characteristics of quantum vortices due to their
experimental versatility \cite{matthews,williams,shaeer,madison,raman}. Besides the studies of static vortices in rapidly rotating trap, optical lattices and dipolar BECs \cite{coopera,cooperb,fetter,feder,jaksch,zhai,zhao,zhangxf}, there have been shown a considerable increase of the interest on the dynamics of vortices. The dynamical properties
of a BEC for vortex nucleation \cite{sinha,jianchun,jacksona,yang} and evolution \cite{caradoc1999a,allen,jacksonb,wrighttm,rooney,yan} have been investigated extensively.
In addition, much work has been done on the vortex structure in multi-component \cite{mueller,kasamasu} and spinor BECs \cite{lovegrove,hanjunghoon,leanhardt}, as well as the collisions and evolution of vortex loops and knots in 3-dimensional BECs \cite{koplik,caradoc2000b,kobayashi,kaneda,borgh,klecknera,klecknerb,hall}.

Since the vortex dipoles have been directly observed in a pancake-shaped condensate by forcing superfluid flow around a repulsive Gaussian obstacle
potential\cite{Freilich,neely}.The topic of excitation about vortex dipoles has recently attracted extensively theoretical and experimental studies,
which mainly concentrate on the vortex shedding mechanism\cite{kadokura,kwona,kwonb,kwonc}, dynamics excitation\cite{kwonb,seo2016a} and persistent current\cite{law2014a}.
Additionally, vortex dipoles induced in oscillating  potential\cite{fujimoto}, at finite temperature\cite{gautam} and under spin-orbit coupling\cite{kato2017a} are
all studied. Now, the technique using two blue-detuned lase beams as"tweezers" provides a broad set of possibilities to pin and manipulate vortices on demand experimently\cite{aioi2011a,gertjerenken,samson2016a}. Recent progress in a toroidal geometry \cite{ramanathan,wrightkca,wrightkcb,jendrzejewski,yakia,yakib,eckel}
has opened a new prospect to study the superfluidity and vortex excitation\cite{wrightkca, white2017a, abadm}. When a barrier rotate sufficiently rapid in an annular BEC, surface mode can be excited and come into the condensate\cite{yakib}. Furthermore, a persistent flow in a toroidal trap can be created by stirring with a rotating weak link\cite{wrightkcb}.
In contrast, despite a variety of researches on dynamical response of a BEC stirred by a rotating laser beam\cite{madison,desbuquois2012a,singh}, the dependence of deep
shape deformation of condensate on rotating frequency is never reported. Moreover, comprehensive study of vortex dipoles inside of the obstacle on their nucleation,
splitting properties and the long-time dynamical behavior are less concerned, which are all of our great interests.

In this paper, by solving the time-dependent dissipative Gross-Pitaevskii (GP) equations, we reveal the shape deformation and long-time dynamical behaviors
of an highly oblate BEC stirred by a rotating laser beam.  The density of the condensate can be rotating nearly rigidly with the same frequency to the laser
beam, with the shape of the condensate deformed heavily due to excitations of many surface waves with relatively lower angular momenta. After a long enough time,
a few vortices are found to be left either trapped in the condensate or pinned by the obstacle. On the other hand, stirring BEC by a rotating blue-detuned laser
beam can cause a vortex dipole or several vortices to be pined at the beam center.

\emph{Model.}---Consider a single-component BEC described by the normalized macroscopic wave function $\psi(\textbf{r}, t)$. In the mean-field framework, the dynamics of a system with $N$ weakly identical atoms close to thermodynamic equilibrium and subject to weak dissipation can be described by the time-dependent GP equation\cite{choi}:
\begin{equation}
(i-\gamma)\hbar\partial_{t}\psi=[\frac{-\hbar^2\nabla^2}{2m}+V(\textbf{r})+Ng|\psi|^2)]\psi,
\end{equation}%
where $V(\textbf{r})=\frac{1}{2}m(\omega_{t}^{2}x^{2}+\omega_{t}^{2}y^{2}+\omega_{z}^{2}z^{2})$ is the axially symmetric harmonic trap potential, and $\omega_t$, $\omega_z$ are the radial and axial trap frequencies. The parameter $\gamma$ in GP equation is a dimensionless, phenomenological, damping constant. It takes into account the quantum and thermal fluctuations from background and is always introduced to fit experiment. The dissipative GP equation has been extensively employed to study the dynamics of systems in the presence of thermally induced dissipation\cite{gertjerenken,samson2016a,yan,yakia,yakib}.

We focus in this paper on a highly oblate BEC with $\omega_z\gg\omega_t$. In this extreme limit, the axial dimension is sufficiently thin that the motion along z direction can be neglected and atoms can move only within the $xy$ plane. The normalized $\psi$ can thus be written as $\psi(\mathbf{r},t)=\psi_{2D}(x,y,t)\phi_{0}(z)$, where $\phi_{0}(z)=(\pi a_{z}^{2})^{-1/4}exp(-z^{2}/2a^{2}_{z})$, with $a_{z}=\sqrt{\hbar/m\omega_z}$ the characteristic length of the longitudinal harmonic oscillator. The atom-atom contact interaction is $g=4\pi\hbar^{2}a_s/m$ with $a_s$ the s-wave scattering length. In the numerical computations, we discretize the x-y plane into a square lattice. The artificial lattice constant $a$ must be much less than the characteristic length $l=\sqrt{\hbar/m\omega_t}$ of the axial harmonic oscillator to validate the simulations. For the time evolution the forth-order Runge-kutta method is employed at each time step. The central-difference formula is used mainly to calculate the kinetic term.
Introducing dimensionless $\psi(i,j)$, by substituting $\psi(\textbf{r})$ with $\frac{1}{\sqrt{a^2a_z}}\psi_(i,j)$, we thus obtain the following lattice-version GP equations:
\begin{equation}
\begin{split}
(i-\gamma)\hbar\partial_{t}\psi(i,j)=\{-t_{0}[\psi(i-1,j)+\psi(i+1,j)\\
+\psi(i,j-1)+\psi(i,j+1)-4\psi(i,j)]\\
+[V(i^2+j^2)+Ng|\psi(i,j)|^2]\psi(i,j)\},
\end{split}
\end{equation}%
where $t_{0}=\frac{\hbar^2}{2ma^2}$, $V=\frac{1}{2}m\omega_t^2a^2$,
$g=\frac{4\pi a_s\hbar^2}{ma^2a_z}$. Since all the parameters $t_{0}$, $V$, $g$ and
$\hbar/t_{0}$ have the scale of energy, it is convenient to introduce
dimensionless parameters $V'=V/t_{0}$, $g'=g/t_{0}$ and
$t'= t/(\hbar/t_{0})$, measured in unit of $t_{0}$.
All of these parameters are actually only dependent of $a/l$ and $a_{s}/a_{z}$. A straightforward analysis leads to the following
expressions of $V'=(\frac{a}{l})^4$ and $g'=8\pi\frac{a_s}{a_z}$. Note that $g'$ is essentially independent
of the artificial lattice constant $a$.

We assume the system consists of $^{87}Rb $ atoms. Thus we have $m\approx87m_p$ with $m_p$ the mass of proton. The trap frequency is chosen to be
$\omega_t=2\pi\times 10 s^{-1}$ and $\omega_z=2\pi\times 100 s^{-1}$, then $l$ is estimated to be about $3.4\mu m$.
When the square lattice we study takes a typical size of $200\times200$ and $(a/l)^{2}$ is chosen to be 0.008, this
means that the system has a size of about $60\mu m\times60\mu m$ and $\hbar\omega_t/t_{0}=2(a/l)^{2}=0.016$, $\hbar/t_{0}=0.2ms$, $V'=0.000064$.
In addition, to guarantee both the convergence and efficiency of iteration of the GP equations, $dt'$ is chosen to be between $10^{-4}$ and $10^{-2}$ in numerical calculations.

\emph{Results.}---When a blue-detuned laser beam is rotating uniformly with angular velocity $\Omega$ in BEC, the obstacle produced by it can be well described as a moving Gaussian potential:
\begin{equation}
V_{GOP}(x,y)=V_{0}exp(\frac{-\{[x-x_{0}(t)]^{2}+[y-y_{0}(t)]^{2}\}}{\sigma^{2}}),
\end{equation}
where $(x_{0}(t),y_{0}(t))=(Rcos\Omega t, Rsin\Omega t)$ with $R$ the distance between the obstacle and trap center, $\sigma$ and $V_{0}$ are the waist and intensity of the laser beam respectively. The detailed behavior of the stirred BEC depends sensitively on $\sigma/\xi$ and $V_{0}$ \cite{aioi2011a,kwona,kwonb,wrightkca,kwonc,inouye}, and in the following they are fixed to be $V_{0}=8t_{0}$, $\sigma=0.8l$. Since the chemical potential is estimated to be $\mu=0.5$, $V_{0}/\mu=16$ means the laser beam we study is a impenetrable obstacle\cite{kwona,kwonb}. By starting from the ground state of the BEC in the presence of a static obstacle, the uncontrollable excitations are prevented from an abrupt motion of the obstacle. Angular velocity of the obstacle is increased linearly and slowly with time until up to the desired value $\Omega$. Then the obstacle moves with a constant angular velocity $\Omega$. In the following, according to the choice of Ref.\cite{yakia} to meet experiments, $\gamma$ is set to be $1.5\times10^{-3}$. We find our results do not depend qualitatively on the specific value of $\gamma$ when it is in the range of 0.001 to 0.005.

\begin{figure}[!htb]
\centering
\includegraphics[scale=1.12]{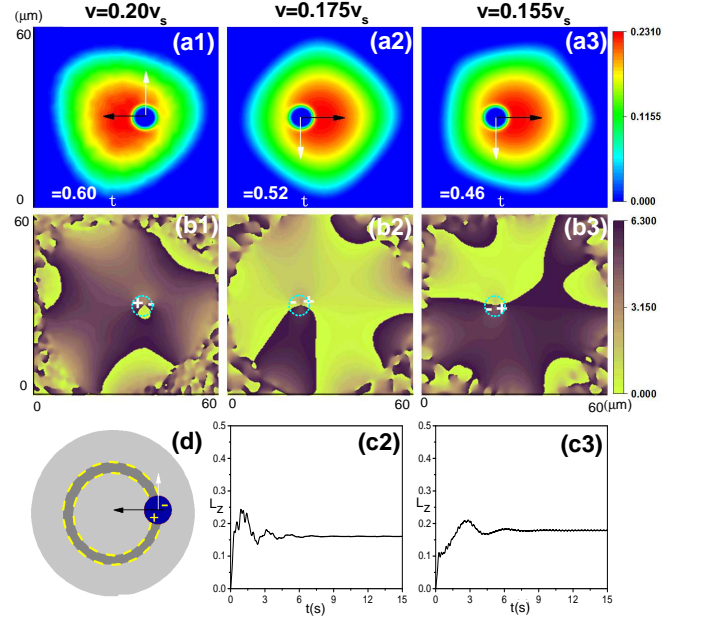}
\caption{(color online) The upper two panels denote the density and phase profiles when the obstacle velocity is below the critical value, where the the generated vortex-antivortex pair has not been split out. Here, $Ng'=5000$. The + and - signs denote the vortex and antivortex of the pair. The white and black arrows show the directions of obstacle motion and vortex dipole, which are always perpendicular to each other. The green dotted circles in (b1),(b2),(b3) show the boundary of the obstacle. The angular momenta per atom $\overline{L}_{z}=-i\hbar\int d\textbf{r}\psi^{*}(\textbf{r})(x\partial y - y\partial x)\psi(\textbf{r})$ as functions of time for (a2) and (a3) are exhibited accordingly in (c2) and (c3). The angular momentum contribution from the vortex dipole is schematically shown in (d), where only the atoms in the dark shaded annular region effectively contribute to $L_{z}$, $\hbar$ per atom. }
\end{figure}

The critical velocity $v_{c}$ is defined as the value of the obstacle velocity $v=R\Omega$ at which the vortex excitation begins to be generated into the BEC by the obstacle. The critical velocity for a moving hard cylinder in an uniform BEC is believed to provide an upper bound for $v_{c}$ \cite{kwona}. Here for a trapped BEC, $v_{c}$ depends not only on the nature of the condensate, but also on the distance $R$. When $R=1.8l$, $v_{c}$ is estimated to be about $v_{c}=0.21v_{s}\mid_{r=R}=242um/s$, where $v_{s}=\sqrt{\frac{ng}{m}}$ is the speed of sound. However even when $v$ is much smaller than $v_{c}$, a vortex-antivortex pair can be produced and trapped by the obstacle. Since the obstacle potential is finite in spite of being large, the local density $n$ of the BEC within the small space occupied by the obstacle is rather small but finite. The small local density thus lowers the speed of sound locally, leading to the local excitations within the obstacle according to Landau criterion. When the obstacle velocity $v$ is below $v_{c}$ but not too small, the vortex dipole direction can be identified from the phase profile of the BEC, which is rotating uniformly within the obstacle but always keep perpendicular to the motion direction of the obstacle, as exhibited in Fig.1.

\begin{figure}[!htb]
\centering
\includegraphics[scale=0.8]{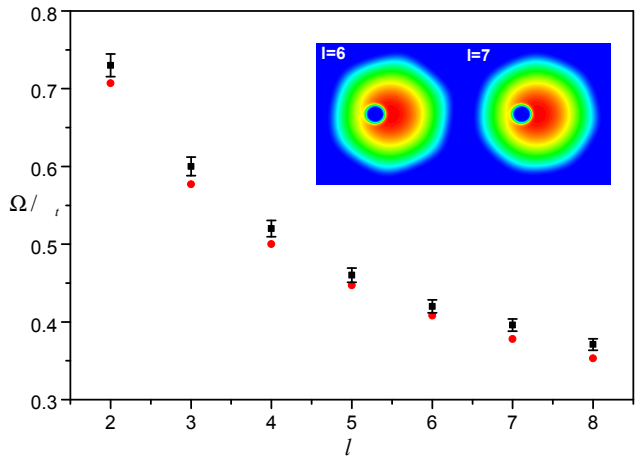}
\caption{(color online) Comparison of the critical frequency $\Omega$ for surface modes excitations as angular momentum $l$ from 2 to 8. The red circles are the analytical values by $\Omega=\omega_{t}/\sqrt{l}$. The black squares are the values obtained numerically and the error bars indicate the standards deviations. The insets are the density profiles of our
numerical results for $l$ being 6 and 7. }
\end{figure}

Besides the critical velocity $v_{c}$, there exists a critical angular velocity $\Omega_{c}=min\{\omega_{l}/l\}$ according to an analog of Landau criterion\cite{onofrio,dalfovo,recati}, where $\omega_{l}$ is the excitation energy for a surface wave. For smaller $l$, $\omega_{l}=\sqrt{l}\omega_{t}$. For the parameters we chose, $\Omega_{c}$ is estimated to be about $0.3\omega_{t}$, above which surface waves can be excited. But the surface waves with larger $l$ can hardly be identified from the density or phase profiles of the BEC obtained numerically. On the other hand, $\omega_{l}/l=\omega_{t}/\sqrt{l}$ can also be viewed as the minimum excitation frequency for generating a surface wave with angular momentum $l$, indicating the surface waves with smaller $l$ can be possibly found upon increasing $\Omega$ up to a sufficient large value. It is found in our calculations that when $v=0.2v_{s}$ which is slightly below $v_{c}$, the density of the BEC is deformed triangularly, as seen in Fig.1(a1), and the BEC is rotating rigidly with the same angular velocity to the obstacle. This is identified as the surface wave with angular momentum $l=3$ because the density difference $\delta n$ due to stirring can be viewed as $\delta n\propto cos(3\theta-\omega_{3}t)$. The motions of the density of the BEC and the obstacle are nearly synchronous, i.e., locked to each other, since $\Omega$ for this case is $\Omega=0.6\omega_{t}\approx\omega_{t}/\sqrt{3}$. The shape of the condensate is deformed heavily, so there should be plenty of surface waves dominated at $l=3$. The existence of many surface-wave excitations can be confirmed by the existence of a finite angular momentum of the condensate(see Fig.1(c2),(c3)), which cannot be interpreted just as the contribution from the vortex dipole. As shown schematically in Fig.1(d), the vortex dipole's contribution to the angular momentum is estimated to be  $\sim\frac{d_{0}}{R_{TF}}\ll0.1$ with $d_{0}$ the vortex-antivortex separation distance and $R_{TF}$ the TF radius, since $d_{0}$ is found to be much less than the healing length $\xi$. Fig.1(a2)((a3)) shows the similar behavior of the condensate with the frequency $\Omega$ slightly larger than $\omega_{t}/2$($\omega_{t}/\sqrt{5}$), where the surface wave excitations are dominated at $l=4$($l=5$), and the condensate is deformed tetragonally(pentagonally), rotating with nearly the same angular velocity as $\Omega$. Since $R\omega_{t}/\sqrt{2}>v_{c}$ for $R=1.8l$, the elliptical deformation due to $l=2$ surface waves cannot be observed. But for a smaller $R=1.3l$, our calculations confirm the existence of the elliptically deformed BEC for $\Omega=\omega_{t}/\sqrt{2}$, which is consistent with the experimental observations\cite{onofrio,madison}. In Fig.2, we plot our results of the critical angular velocity for surface wave excitations as angular momentum $l$ from $2$ to $8$ and comparing them with the analytical value of $\omega_{t}/\sqrt{l}$.
\begin{figure}[!htb]
\centering
\includegraphics[scale=0.75]{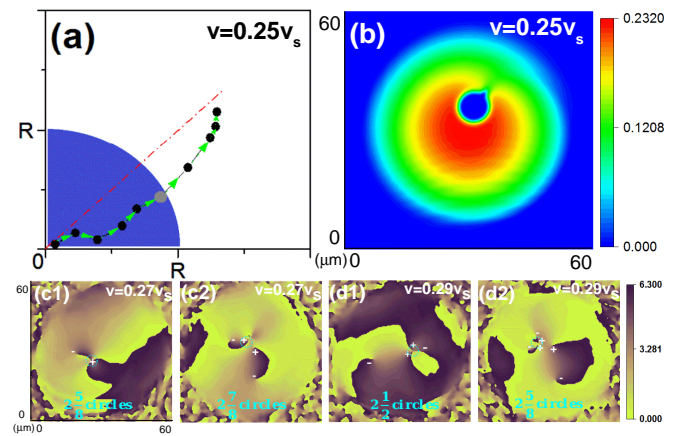}
\caption{(color online) (a) The orbit of the antivortex of the vortex dipole in the frame of reference of the moving obstacle, when the obstacle velocity is slightly above the critical value. Here the blue region represents the density hole of obstacle with radius $R=3.5\xi$ with $\xi$ the healing length at the trap center. The red dash-dotted line denotes that between the centers of the obstacle and trap. The ten circle dots are representative positions of antivortex as it is escaping from the obstacle within the first one half circle of stirring, where the arrows indicate the motion directions of the antivortex. (b) The density profile of the BEC when the antivortex moves to the grey-shaded dot in (a), where the spike denotes the exact position where the antivortex leaves the obstacle. c(1)(d(1)) and c(2)(d(2)) are the phase profiles before and after the generation of the second(third) vortex dipole within one fourth(eighth) circle of stirring when $v=0.27v_{s}$($v=0.29v_{s}$), where only two(three) vortex dipoles are generated in the whole stirring process. }
\end{figure}
Once the obstacle velocity $v$ is reaching $v_{c}$, the antivortex starts to separate from the vortex and move towards the barrier edge, while the vortex is still pinned at the obstacle center, as schematically shown in Fig.3(a). Meanwhile, a density spike is formed (Fig.3(b)), long before the antivortex leaves the barrier edge. The antivotex is then released from the obstacle and moves gradually to the boundary of the condensate. When $v$ is slightly above $0.26v_{s}$, after the first antivortex escapes from the obstacle, another vortex-antivortex pair will be generated and subsequently the new antivortex starts to escape, leaving the two vortices trapped at center by the obstacle(see Fig.3).
When $v>0.3v_{s}$, more vortex-antivortex pairs will be generated during the whole stirring process and more vortices can be trapped by the obstacle.
\begin{figure}[!htb]
\centering
\includegraphics[scale=1.045]{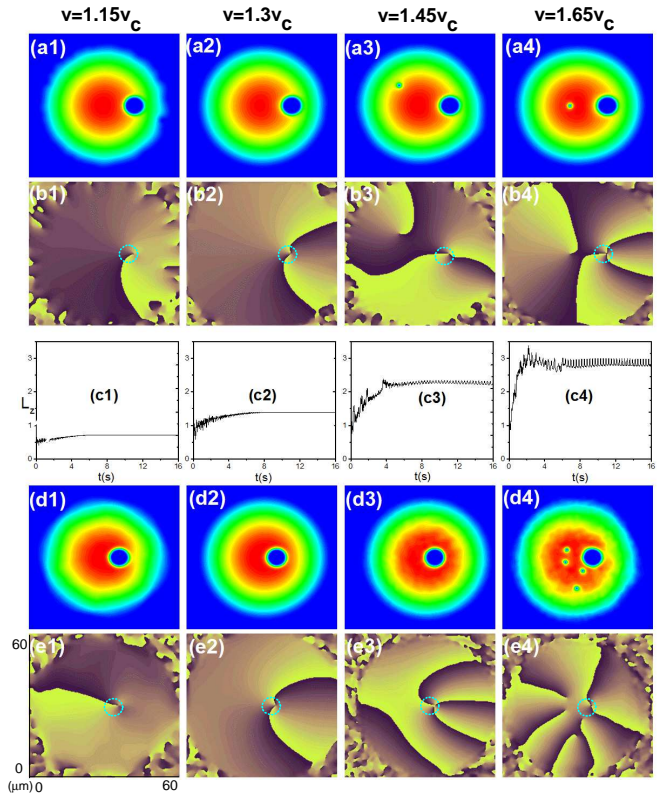}
\caption{(color online) The long-time dynamical behavior of the condensate when the obstacle velocity is above the critical value $v_{c}$ with $v_{c}=255um/s$ at $R=3.6l$ and $v_{c}=242um/s$ at $R=1.8l$, by stirring the BEC for over one hundred circles. The upper(lower) two panels denote the density and phase profiles with $R=3.6l$($R=1.8l$). The middle panels denote the angular momenta per atom corresponding to the cases from (a1) to (a4) respectively. }
\end{figure}

After stirring for over one hundred circles, the state of the BEC becomes dynamical stable and shows the nearly time-independent features. This means the obstacle will stop shedding vortices after a long-time stirring. This can be better understood in the limit of $V_{0}$ being infinite. The vortices trapped by the obstacle or the condensate have changed the distribution of the superfluid velocity around the obstacle. According to Landau criterion, no vortex will be excited, if all velocities are below the critical value, leading to a steady persistent current state after a long-time stirring. Now we study these long-time behaviors of the BEC and focus on the vortex number left in the condensate. After long enough time, all antivortices will leave the condensate and only vortices are left. Among the remaining vortices, several are pinned by the obstacle, the others are loosely trapped near the trap center. The nearly time-independent angular momentum also confirms the stability of the long-time behavior of the BEC. In the parameter region we studied, three vortices at most can be trapped by the obstacle. These results are demonstrated in Fig.4. The number of trapped vortices is summarized in Table 1. If the laser beam stops moving suddenly, the trapped vortices will be released but finally a persistent current state with one or two vortices trapped at the obstacle center is found to be stabilized. When the laser beam is ramping off, the trapped vortices will become free and will then move to the boundary. These make the vortex manipulation by laser beam possible. For the parameters we studied here we only focus on the velocity regime $v<1.7v_{c}$, since when $v>2v_{c}$, so many vortices are generated quickly that the stirred BEC will be in a turbulent state.
Within the regime $v_{c}<v<1.7v_{c}$, it is not from the boundary but within the obstacle center in the form of vortex pair that the vortices are created.
\begin{table}
\centering
%\fontsize{6}{7}\selectfont
%\begin{threeparttable}
\caption{The number of trapped vortices after stirring BEC for a long time, for the parameters we studied.
While $n_{1}$ denotes the number of trapped vortices within the obstacle, $n_{2}$ denotes that
outside the obstacle. The total number is $n=n_{1}+n_{2}$.}
\label{tab:1}
%\scalebox{1.2}{
\begin{tabular}{c r r r r r r}
%\toprule
\hline
\hline
\multirow{2}{*}{$v$}&\multicolumn{3}{c}{\hspace{0.33cm}R=3.6$l$}&\multicolumn{3}{c}{\hspace{0.45cm}R=1.8$l$}\cr
\cline{2-7}
%\cmidrule(lr){2-4}\hspace{0.22cm}\cmidrule(lr){5-7}
&\hspace{0.4cm}$n_{1}$&\hspace{0.4cm}$n_{2}$&\hspace{0.4cm}$n$&\hspace{0.5cm}$n_{1}$&\hspace{0.4cm}$n_{2}$&\hspace{0.4cm}$n$\cr
\hline
%\midrule
1.00$v_{c}$--1.25$v_{c}$&1&0&1&1&0&1\cr\hline
1.25$v_{c}$--1.35$v_{c}$&2&0&2&2&0&2\cr\hline
1.35$v_{c}$--1.50$v_{c}$&2&1&3&3&0&3\cr\hline
1.50$v_{c}$--1.70$v_{c}$&3&1&4&2&4&6\cr\hline
\hline
%\bottomrule
\end{tabular}
\end{table}

\emph{Conclusion.}---We have performed the numerical calculations of the quasi-two-dimensional GP equation to investigate the surface wave excitations and long-time behavior of vortices in a stirred highly oblate condensate. Plenty of surface waves can be excited with the condensate shape being deformed heavily according to the rotation frequency. Vortex-antivortex pair can be created in the obstacle center even when the obstacle velocity is relatively smaller. Once the obstacle velocity reaches a critical value, the antivortex starts to separate from the vortex and then leaves the obstacle regime. Furthermore, after a long enough time, a few vortices are found to be left either trapped in the condensate or pinned by the obstacle and the number of them depends on the velocity and position of the obstacle.
\begin{acknowledgments}
This work is supported by the NSFC Project No.11174126 and 973 Projects No.2015CB921202.
\end{acknowledgments}

\bibliography{ref}

\end{document}